# DARK MATTER IN GLOBULAR CLUSTERS

DOUGLAS C. HEGGIE
*University of Edinburgh*
*Department of Mathematics and Statistics, King's Buildings,*
*Edinburgh EH9 3JZ, U.K.*

AND

PIET HUT
*Institute for Advanced Study*
*Princeton, New Jersey 08540, U.S.A.*

**Abstract.** We first review reasons why dark matter is an interesting issue in connection with star clusters. Next we consider to what extent the presence of dark matter is consistent with their dynamics and structure. We review various model-dependent and model-independent methods which have been applied to two well studied clusters, NGC 6397 and 47 Tuc. We suggest that about half of the mass in each object is still unobserved, possibly in the form of a mixture of low-mass stars and white dwarfs.

## 1. Introduction

It is remarkable how many of the major problems in astronomy are influenced by what we know about globular clusters. Nevertheless, the dark matter problem is an exception. The question whether globular clusters contain significant amounts of dark matter has not often emerged in the literature, however much it may have been discussed informally, and to the best of our knowledge it has never previously been reviewed.

In this paper we first attempt to set out the background to the problem, explaining why the search for dark matter in the context of globular clusters is an interesting problem. Next we consider ways in which such a search may be carried out, with particular emphasis on techniques based on dynamical modelling. We shall see that this topic has reached a new and exciting point of development, thanks to the depth to which the mass function of globular clusters can now be examined observationally. Nevertheless important uncertainties still remain, and direct attention to both theoretical and observational problems for the coming years.

## 2. The Importance of Dark Matter in Globular Clusters

There are several *a priori* reasons why the search for dark matter may be taken seriously.

1. In terms of mass, globular clusters are the next step down from the smallest stellar systems which definitely contain dark matter – the dwarf spheroidals (see Ashman 1992, Pryor 1992).
2. The presence or absence of dark matter in globulars would be a significant piece of evidence in the study of galaxy formation. It would clarify the relation between globulars and their host galaxy, and between globulars and other small stellar systems.
3. Renewed interest in the topic is timely because of the flood of new results on deep mass functions, and the observation of white dwarfs, in several globular clusters (cf. the papers by Piotto and by Fahlman, these proceedings).
4. Hypothetical black hole stellar remnants were invoked by Larson (1984) to account for anomalously high-velocity stars in some clusters.
5. Upper limits on the dark matter content of globulars are needed in order to assess the feasibility of searches by such techniques as microlensing (Griest, pers. comm.).
6. Though the issue seems controversial, there is a possibility that certain types of dark matter might affect the evolution of stars (Renzini 1987, and references in Heggie *et al.* 1993).
7. The typical masses of globular clusters have a special significance in reasonably standard cosmological models (Peebles 1984, Rosenblatt *et al.* 1988, West 1993), and Peebles suggested that globulars might contain a reasonably uniform dark matter component. Other theories (e.g. Silk & Stebbins 1993, Moore & Silk 1995) are also relevant to the structure and content of globular clusters, but give a less clear picture of the expected distribution of the dark matter.

## 3. Searching for Dark Matter in Globular Clusters

### 3.1. SEARCH STRATEGIES

How one searches for dark matter depends on what it is, and in principle one may consider all the usual suspects: black holes, neutron stars, white dwarfs, low-mass stars, wimps, etc (see Carr 1994 for a review, especially on baryonic varieties). Certain types of dark matter could be detected from the resulting gamma-ray emission (cf. references in Heggie *et al.* 1993). On the more conventional side, low-mass stars and brown dwarfs may be detected by a variety of means. Detection and spectroscopy from the ground in the IR may be feasible with 8m class telescopes (Fusi Pecci *et al.* 1994). For estimates of the usefulness of space-based observations, and of microlensing, see the paper by Longaretti *et al.* in this volume, and references therein.

In the remainder of this paper we consider the traditional *model-building strategy*. It is the counterpart in globular clusters of the classical Oort technique which revealed the possibility of missing mass in the solar neighbourhood (Oort 1932). In principle one should construct a model of a cluster which is consistent with all relevant dynamical data, including (i) the surface brightness profile; (ii) star counts; (iii) radial velocities; (iv) proper motions; (v) pulsar "spin-up" (see the paper by Kulkarni in these proceedings); and (vi) dynamical evolution. Finally the mass of the model may be compared with that of visible matter in order to determine the amount of dark matter.

In practice dynamical models are often constructed from a subset of the above list of data. For example, the commonest models are constructed from the surface brightness profile only. Without kinematic data, however, these do not constrain the mass sufficiently. (The most obvious illustration of this assertion is a single-component model, to which any amount of dark stars of the same mass could be added without altering the surface brightness profile; see also Longaretti *et al.*, this volume.) Though radial velocity data is rather commonly used, it is unusual for data on proper motions to be taken into account, Leonard *et al.* (1992) being a notable exception.

An example of a cluster in which this technique has revealed the presence of dark matter is M71. Richer & Fahlman (1989) obtained a mass-to-light ratio $(M/L)_V \simeq 0.57$ from star counts out to radius $3.4'$, compared with a global value in the range $1 - 1.4$ obtained by dynamical modelling of kinematic data. This result indicates that at least about 50% of the mass of this cluster resides in unobserved components, and Richer & Fahlman concluded, after further modelling, that it was most likely to consist of stars with masses below that of the faintest stars observable in their study, i.e. less than about $0.33 M_\odot$.

What makes a reexamination of such investigations timely is that it now seems possible to push star counts down to masses closer to $0.1 M_\odot$ in some clusters, as the following examples illustrate. And another dark matter candidate which can now be counted convincingly is the population of white dwarfs, or at least the brightest ones. It has already been shown (see the paper by Fahlman, these proceedings) that their numbers correspond nearly to what would be expected from the evolution of stars originally slightly more massive than the present turnoff. Therefore they may be expected to contribute substantially to any missing mass.

The following two examples are meant to illustrate the advances in modelling to which these new observational data should lead. In addition, however, it is our aim to illustrate the variety of modelling techniques which are now available. We shall try to show that each has its strengths and weaknesses, and that use of several techniques for the same cluster helps to guard against the risk of drawing model-dependent conclusions.

3.2. EXAMPLE: NGC 6397

As the frequency of its appearance in the papers in this volume show, this cluster has almost replaced M15 as the classic example of a post-collapse cluster, one reason being its relative proximity. Like several post-collapse clusters, it exhibits population gradients (Djorgovski *et al.* 1991). According to Aguilar *et al.* (1988) it is one of the most fragile galactic globular clusters.

Anisotropic multi-mass King models for NGC 6397 can be found in Meylan & Mayor (1991), who fitted to radial velocities and the surface brightness profile. An isotropic model was constructed by King *et al.* (1995), who fitted to star counts and the surface brightness profile, but no kinematic data. This last study furnishes a beautiful example of mass segregation in observational data, and the fit to the surface brightness profile is excellent. There is also a single-component King model given by Da Costa (1979).

As already mentioned, deep star counts and kinematic data are important for establishing the existence of dark matter, and so we have constructed a multi-mass isotropic King model which fits the projected radial velocity dispersion profile of Meylan & Mayor (their Table 2), as well as their $V$-band surface brightness profile and the deep star counts of Paresce *et al.* (1995a). (Note, incidentally, that there exists some disagreement between different groups [King, pers. comm., and Piotto, this volume] in the counts of the stars of lowest mass.) We used the same mass bins as Meylan & Mayor, and our best fitting model has the global mass function given in Table 1. Other parameters of the model, in standard notation, are $W_0 = 12.7$, $\sigma^2 \equiv 1/(2j^2) = 12.0 \text{km}^2\text{s}^{-2}$, $r_c = 0.19\text{pc}$, and $r_t = 26\text{pc}$. We have checked that the resulting model is consistent with the star counts of King *et al.* at two radii. The fit to the surface brightness profile is of comparable quality (judged by $\chi^2$) to that of the models of Meylan & Mayor.

At this point it is worth pausing to consider the limitations and advantages of multi-mass King models. Their principal advantage (and it is a formidable one, which explains their great popularity) is speed. Among their limitations, however, are

- *Neglect of anisotropy, rotation, flattening* All three pose considerable

TABLE 1. Global Mass Function for a New Model of NGC 6397

| Mass range ($M_\odot$) | Mean mass ($M_\odot$) | Mass ($M_\odot$) |
| --- | --- | --- |
| $[5, 100]_{hr}$ | 1.400 | 400 |
| $[2.5, 5]_{wd}$ | 1.093 | 5600 |
| $[1.5, 2.5]_{wd} + [0.63, 0.79]$ | 0.726 | 3400 + 9600 |
| $[0.79, 1.5]_{wd} + [0.50, 0.63]$ | 0.578 | 23700 + 5500 |
| $[0.40, 0.50]$ | 0.447 | 3400 |
| $[0.32, 0.40]$ | 0.355 | 2900 |
| $[0.25, 0.32]$ | 0.282 | 3400 |
| $[0.20, 0.25]$ | 0.224 | 2300 |
| $[0.16, 0.20]$ | 0.178 | 1800 |
| $[0.13, 0.16]$ | 0.141 | 1000 |
|  | Total cluster mass | 63000 |

Note: the notation for the mass range follows that of Meylan & Mayor exactly. Thus the first bin contains heavy remnants of stars with initial mass in the stated range. Similarly *wd* denotes white dwarf remnants. Other ranges refer to main sequence stars.

modelling problems, and even though anisotropy is often included, it is far from clear on dynamical grounds that the usual recipe (i.e. Michie-King models) is at all appropriate.
- *Approximate dynamical evolution* The lowered Maxwellian distribution was introduced as an approximate solution of the one-component Fokker-Planck equation. For a long time, however, it has been possible to solve this equation by direct numerical methods (see below).
- *Lack of primordial binaries* These effectively give rise to a small population of bright objects more massive than the turnoff mass. They are usually ignored, though the work of King *et al.* is a notable exception.
- *Problems of population gradients* The implication is that the surface brightness profile is sampling different populations at different radii, and in principle this may affect modelling based on the surface brightness profile.
- *Poor statistical methodology* It has been claimed (Merritt & Tremblay 1994) that astronomy is one of the last disciplines to hold out against the trend towards non-parametric statistics, and the use of parametrised models such as multi-mass King models introduces unquantified biases and other undesirable deficiencies.

Many of these drawbacks can be put right, but usually at the cost of computational speed. Dynamical evolution can be handled better by means of Fokker-Planck calculations, and in fact Drukier (1994) has carried out an excellent study of this cluster using this technique. To some extent his preferred models were guided by ground-based faint star counts which have now been superseded, but there is little doubt that modest adjustment of the parameters of his models would be sufficient to restore good agreement.

The limitations of the Fokker-Planck method include

- *Time-consuming computations*
- *Neglect of anisotropy, rotation and flattening*, though it is now becoming possible to include anisotropy efficiently (see the papers by Takahashi and Giersz in these proceedings). Also there has been a modest recent revival of interest in the Fokker-Planck modelling of rotating clusters.
- *Omission of disk shocking*, though this could easily be included at a satisfactory level of approximation.
- *Omission of primordial binaries*, despite their known importance in core bounce and post-collapse evolution (see Hut, these proceedings). They could be included, but the necessary cross sections are not well known, especially for unequal masses, and the level of approximation would be rough. Monte Carlo methods (see Giersz, these proceedings) offer the best prospect here.

TABLE 2. Inferred data for NGC 6397

| Source | Tidal Radius pc | Total Mass $10^5 M_\odot$ | Neutron Stars % by mass | White Dwarfs % by mass |
|---|---|---|---|---|
| Da Costa (1979) | 24 | – | – | – |
| Meylan & Mayor (1991) | $66 \pm 14$ | $1 \pm 0.1$ | 2 | 25 |
| Drukier (1994) | $19 \pm 2$ | $0.66 \pm 0.05$ | 4 | ? |
| This paper | 26 | 0.63 | 0.6 | 52 |

Note: where necessary we have assumed $1\text{pc} = 1'.6$.

To return to NGC 6397, Table 2 summarises the main data from the various models which are relevant to the dark matter problem, though insufficient details of the model of King *et al.* were available to us. A comparison between the data derived from different models or different selections of observational data helps to assess the the magnitude of the systematic errors in these estimations. The high mass of the model of Meylan & Mayor, for example, may stem from their use of anisotropy, which extends the halo. In any case $r_t$ is difficult to determine for this cluster because of the high background density (Drukier *et al.* 1993).

The interesting number in this table is the proportion of white dwarfs, which we think may be higher than was previously believed. Our model also contains a lower proportion of low-mass stars (in the last six bins in Table 1) than in the models of Meylan & Mayor. At the time when the latter models were constructed the main sequence mass function was poorly constrained.

### 3.3. EXAMPLE: 47 TUC

It can be claimed that conclusions such as this are still too model-dependent, and could be relaxed if some of the specific choices of multi-mass King models (e.g. the choice of distribution function) were altered. One of the aims of non-parametric methods is to avoid this pitfall. Of the four clusters studied non-parametrically by Gebhardt & Fischer (1995), we select 47 Tuc, being one of those for which recent deep star counts have become available. Though it would be desirable to construct new King models taking account of this data, we have not yet done so, and it is interesting to see what conclusions can be drawn from the above non-parametric study alone.

The dynamical status of 47 Tuc is a little controversial, though it is commonly assumed to be a high-concentration cluster approaching core collapse. Without doubt it is amongst the most massive clusters.

The advantages of non-parametric methods have already been touched upon, and it is worth listing their possible defects. These include

- *Neglect of anisotropy, rotation and flattening*
- *Neglect of dynamical and physical constraints* The method makes no assumption about the form of distribution function of the population used to trace the potential, even where it may be well constrained by theory. The method also makes no assumption that the mass density is positive, and in fact the results inferred for 47 Tuc imply that a negative mass-to-light ratio is acceptable at the 90% confidence level in one range of radius. This defect may be related with the previous item (concerning isotropy), as Richstone has pointed out (pers. comm.) that the problem with $M/L$ is avoided if some anisotropy is introduced.
- *Effect of population gradients.* In fact Guhathakurta *et al.* (1992) have reported the existence of a population gradient in 47 Tuc. As with multi-mass King models fitted to the surface brightness, however, it is not known how important the effect may be.

From the results of Gebhardt & Fischer we compute that the mass within a sphere of $7'$ projected radius is about $6.7\times 10^5\,M_\odot$. This is close to the value of approximately $5.5\times 10^5\,M_\odot$ inferred from multi-mass anisotropic King-Michie models fitted by Meylan (1989). The result just mentioned suggests that the differences between these methods may be rather philosophical than substantive. Indeed, one can think of the model-fitting method of Meylan as simply a different way of constructing a rather arbitrary potential. Only the mass bin containing the giants is directly connected to the observations; the others simply give rise to a potential field sufficient to agree with the kinematic data. The heaviest bins govern the potential at small radii, and successive bins build up the potential well at successively greater radii.

Now let us consider the surface density at $4.6'$, where deep star counts were obtained by De Marchi & Paresce (1995). By summing their mass function and taking into account the field area, we obtained a surface density in counted main sequence stars of about $305 M_\odot/\mathrm{pc}^2$. This may be compared with the projected density of all matter which we computed from the non-parametric model of Gebhardt & Fischer; this value is approximately $1100 M_\odot/\mathrm{pc}^2$, with a lower limit of about $770 M_\odot/\mathrm{pc}^2$ at 90% confidence.

At face value these data imply a substantial fraction of "missing mass", and we immediately mention some possible explanations.

- *Giants* These were excluded from the counts of De Marchi & Paresce. The mean mass of the stars in their most massive bin was about $0.75 M_\odot$, and we estimate that inclusion of stars between this bin and turnoff ($\sim 0.9 M_\odot$, cf. Hesser *et al.* 1987) would increase the surface density to about $385 M_\odot/\mathrm{pc}^2$. This implies that the proportion of the projected mass unaccounted for is still at least 50%, and it could be as high as about 70% (Table 3).
- *Low-mass stars* All the remaining mass could be accounted for if, below the least massive stars counted, the mass function varies as $dN(m) \propto m^{-1-x}dm$ with $x \simeq 0.6$. Though De Marchi & Paresce find that the mass function flattens at low masses, this conclusion is controversial (see the paper by Piotto, this volume).
- *M/L relation* This is controversial for stars of low mass, and errors here can lead to large differences in the inferred mass function. Note, however, that the surface density is obtained from the *magnitude* distribution $N(m)$ by the integral $\int M(m)dN(m)$, where $M$ is the mass of a star of magnitude $m$. Several $M/L$ relations are plotted by De Marchi & Paresce, and indicate that the resulting uncertainty in the projected mass density is at most 0.1 dex.
- *Completeness* De Marchi & Paresce claim that, even at the faintest bin, their counts are at least 67% complete. This has been corrected

for in the data which we used to compute the projected density.
– *White dwarfs* Previous estimates of the mass fraction in all white dwarfs are given in Table 3.

TABLE 3. Mass fraction of white dwarfs in 47 Tuc

| Source | Mass fraction (%) |
|---|---|
| Illingworth & King 1977 | 31 |
| Da Costa & Freeman 1985 | 35 |
| Meylan & Mayor 1986 | 12–35 |
| Meylan 1988 | 27–37 |
| Meylan 1989 | 16–23 |
| This paper | $\lesssim 70$ |

Though the first five are global estimates, it seems unlikely that at the radius of these observations (about 0.7 half-mass radii) mass segregation could produce a much larger proportion of white dwarfs. Now that white dwarfs can be counted in 47 Tuc (Paresce *et al.* 1995b) it is worth reexamining the sorts of models listed in Table 3 to check whether the population of white dwarfs could not perhaps be rather more significant than was previously thought. The result could also illuminate the still rather controversial evidence on the presence of cataclysmic variables in clusters such as 47 Tuc.

## 4. Conclusions

We wish to emphasise that all kinds of dynamical models are useful in this kind of study. All have some advantages, but the list of their limitations is depressingly long. Studying the same cluster by different techniques is an important way of guarding against some of these.

Based on such methods, the studies of the two clusters on which we have concentrated suggest that a large fraction of their mass, around 50%, is still unobserved. It is not implausible, however, that all of this can be accounted for by white dwarfs or low-mass stars, and so we consider that this represents a generous upper limit on the mass fraction of other, more exotic forms of dark matter in these two clusters. In other words, though there are several good reasons for studying the dark matter problem in globular clusters, it is first necessary to improve our knowledge of the abundance of white dwarfs and low-mass stars. At present it is not even clear which of these might dominate. But the time is ripe for renewed study of these low-luminosity components, thanks to the wealth of new observational data.

## 5. Acknowledgements

We thank several participants of the Symposium, too numerous to mention individually, who commented on our paper during the question session and also privately. We are grateful also to Ivan King and Pierre-Yves Longaretti for their comments on an earlier draft of this written version. We have tried to incorporate all these remarks in the present version.